\documentclass[amsmath,notitlepage,amssymb,aps,showkeys,floatfix,prd,a4paper,
  twocolumn,nofootinbib]{revtex4-1}

\usepackage{graphicx}
\usepackage{amsmath}
\usepackage{epstopdf}
\usepackage{amsfonts,amssymb}
\usepackage{float}
\usepackage[utf8]{inputenc}
\usepackage{pstricks}
\usepackage{color}
\usepackage[compat=1.0.0]{tikz-feynman}
\usepackage{simplewick}
\usepackage{color, colortbl}
\definecolor{Gray}{gray}{0.9}
\usepackage{float}
\usepackage{tikz-feynman}
\usepackage{feynmp}
\usepackage{forest}

\usepackage{verbatim}    
\usepackage{dcolumn}
\usepackage{bm}
\usepackage{epsfig}
\usepackage{braket}
\usepackage[normalem]{ulem} 
\usepackage{longtable,booktabs}
\usepackage{booktabs}

\newcommand{\be}{\begin{equation}}
\newcommand{\ee}{\end{equation}}
\newcommand{\ben}{\begin{eqnarray}}
\newcommand{\een}{\end{eqnarray}}

\def\MeV{\mbox{ MeV}}

\def\mb{\mbox{ mb}} 
\newcommand{\pslash}{\not{\hbox{\kern-2.3pt $p$}}}
\newcommand{\pdslash}{\not{\hbox{\kern-2pt $\partial$}}}

\begin{document}

\title{The $X(3872)$ to $\psi (2S)$ yield ratio in heavy-ion collisions }

\author{ L. M. Abreu$^{1,2}$} \email{luciano.abreu@ufba.br}
\author{F. S. Navarra$^{2}$} \email{navarra@if.usp.br}
\author{H. P. L. Vieira$^{1}$} \email{hildeson.paulo@ufba.br}

\affiliation{$^{1}$Instituto de F\'isica, Universidade Federal da Bahia,
Campus Universit\'ario de Ondina, 40170-115, Bahia, Brazil}

\affiliation{$^{2}$Instituto de F\'{\i}sica, Universidade de S\~{a}o Paulo, 
Rua do Mat\~ao, 1371, CEP 05508-090,  S\~{a}o Paulo, SP, Brazil}


\begin{abstract}

In this work we evaluate the $X(3872)$ to  $ \psi (2S) $ yield ratio            
($N_X/N_{\psi(2S)}$) in  Pb Pb collisions, taking into account the interactions 
of the $\psi (2S) $ and $ X(3872)$ states with light mesons in the  hadron gas 
formed at the late stages of these collisions.  
We employ an effective Lagrangian approach to estimate the thermally-averaged 
cross sections for the production and absorption  of the $\psi(2S)$  and use  
them in the rate equation to determine the time evolution of  $N_{\psi(2S)}$.   
The multiplicity of these states at the end of mixed phase is obtained from the 
coalescence model. The multiplicity of $X(3872)$, treated as a bound state of  
$(D\bar D^{*} +c.c.)$ and also as a compact tetraquark, was already calculated 
in previous works.  Knowing these yields,  we derive predictions for the ratio 
($N_X/N_{\psi(2S)}$) as a function of the centrality, of the center-of-mass 
energy and of the charged hadron multiplicity measured at midrapidity          
$[dN_{ch}/d\eta \,(\eta<0.5)]$. Finally, we make predictions for this ratio in 
Pb Pb collisions at $\sqrt{s_{NN}} = 5.02$ TeV to be measured by the ALICE 
Collaboration in the Run 3.

\end{abstract}

\date{\today}


\maketitle

\section{Introduction}
\label{Introduction}

Among the new hadrons observed in the last two decades~\cite{Workman:2022ynf}, several have properties incompatible with the quark model predictions and can be classified as unconventional states. For recent reviews see~\cite{yuan22,nora,zhu}. Their properties remain subject  
of an intense debate and this makes the exotic quarkonium spectroscopy a hot topic of research. The most relevant question is: which is their structure?  They can be weakly bound hadron molecules, compact multiquark states, cusps generated from kinematical singularities,  excited conventional hadrons, glueballs, hybrids, etc., or even a superposition of different configurations! So far there is no compelling answer to this question. 

One emblematic example is the first-observed and most famous exotic state, the $X(3872)$~\cite{Workman:2022ynf,Belle:2003nnu}. Its intrinsic nature is still matter of controversy. Two configurations are the most explored in the literature~\cite{yuan22,nora,zhu}:  the shallow bound state of open charm mesons $(D\bar D^{*} +c.c.)$ and the $c \bar c q \bar q $ compact tetraquark. The ultimate goal of the present work is to contribute to the 
determination of the  $X(3872)$ structure. 

A new era in the investigation of exotic charmonium states has started with the first observation of $X(3872)$ in relativistic heavy ion collisions reported recently by the CMS Collaboration~\cite{CMS:2021znk}. The data were collected in lead-lead (Pb Pb) collisions at a 
center-of-mass energy $\sqrt{s}=5.02$ TeV per nucleon pair, using the decay chain 
$X(3872) \to J/\psi \pi^+ \pi^- \to  \mu^+ \mu^- \pi^+ \pi^- $. The rapidity and transverse momentum intervals considered were $|y| < 1.6 $ 
and $15 <  p_T < 50 $ GeV.  
The significance of the inclusive $X(3872)$ signal was $4.2 \, \sigma$. Interestingly, the  
prompt $X(3872)$ to $\psi (2S)$ yield ratio  was found to be: 
\begin{eqnarray}
 \mathcal{R} & = & \frac{N_{X(3872)}}{N_{\psi(2S)}} \nonumber \\
    & = & 1.08 \pm 0.49(\mathrm{stat}) \pm 0.52(\mathrm{syst}). 
\label{ratioCMSExp}
\end{eqnarray}
This central value is about one order of magnitude higher than the one observed in $pp$ collisions \cite{lhcb1}, which is close to $0.09$. 
Additionally, the LHCb Collaboration reported the observation of  $X(3872)$ in $p$ Pb collisions at both forward and backward rapidity and at $\sqrt{s}=8.16$ TeV per 
nucleon pair~\cite{LHCb:2022ixc}. The decay chain was the same as the one studied by the CMS Collaboration. The transverse momentum interval considered was $p_T > 5 $ GeV, and the rapidity intervals were $1.5<y<4$ for $p$ Pb (forward configuration) and   
$-5<y<-2$ for  Pb $p$ (backward configuration). The resulting ratios of the $X(3872)$ to $\psi(2S)$ multiplicities  were $0.27 \pm 0.08 \pm 0.05$ in $p$ Pb and $0.36 \pm 0.15 \pm 0.11$ in Pb $p$. Both are bigger than the one seen in $pp$ collisions but smaller than the one observed in  Pb Pb collisions.

At first sight, the comparison between the ratios measured in pp and Pb Pb collisions shows a growth that we might be tempted to 
interpret as some medium effect. However the $p_T$ values measured by the CMS are so large that it is unlikely that the particles carrying these momenta have a thermal and/or hydrodynamical origin. Rather, they are produced perturbatively at the very beginning of the collision and do not
experience any thermalization or interactions with a medium.

On the theoretical side, some recent works have attempted to describe the data. In Ref.~\cite{Yun:2022evm}  the coalescence model has been   
used to estimate $N_X$ and the statistical hadronization model to obtain $N_{\Psi(2S)}$. The obtained ratio has a central value of 0.806 for 
the $X(3872)$ in the molecular configuration and 0.204 for the  tetraquark configuration. In Ref.~\cite{Yun:2022evm} the interaction of 
$X$ (and also of  the $\psi(2S)$) with the light mesons in the hadron gas was not considered. Here we try to estimate the effect of these interactions on the ratio $\mathcal{R}$.   

 At the beginning of a heavy-ion collision  quark-gluon plasma (QGP) is formed. It expands, cools down and hadronizes into a hot hadron gas. 
 The gas lives for about $10$ fm and freezes out generating the observed particles. Conventional and exotic hadrons formed in the end of the mixed phase  can interact with the (mostly light) particles in the hadron gas and their multiplicities may experience modifications due to production and absorption processes, as was pointed out in previous works~\cite{Houng:2013,Torres:2014,abreu16,Abreu:2018mnc,Abreu:2021,Abreu:2022EPJC,QCDSR:21,Abreu:2022jmi,      
 Navarra:2022,Abreu:2023aqy,Abreu:2023awj}. The case of the $X(3872)$ has been studied~\cite{Houng:2013,Torres:2014,
 abreu16,QCDSR:21} and its final multiplicity will depend on the interaction cross sections, which, in turn, depend on the spatial 
 configuration of the quarks. Meson molecules are larger, and therefore have greater cross sections and  stronger interaction with the 
 hadronic medium than compact tetraquarks. In order to have a more complete description of the process these interactions should be taken 
 into account.

In this work we evaluate the  ratio $\mathcal{R}$ 
taking into account the interactions of the $\psi (2S) $ and of the $ X(3872)$ states with the hadron gas formed in heavy ion collisions. We will make use of effective Lagrangians to estimate the thermally averaged cross sections for  $X(3872)$ and $ \psi (2S) $ production and  absorption and employ them in the rate equation to determine the time evolution of the  ratio $\mathcal{R}$.  We will use the coalescence model to
compute the multiplicity of these states at the end of mixed phase.   The $X(3872)$ will be treated as a bound state of $(D\bar D^{*} +c.c.)$  and also as a 
compact tetraquark. Finally, we will make predictions for $\mathcal{R}$ to be 
measured by the 
ALICE Collaboration in the Run 3.

\section{Effective formalism}
\label{Framework} 

In what follows  we describe the effective formalism used to evaluate the 
interactions of the $X(3872)$ and $\psi (2S)$ states with the surrounding  
hadronic medium. In particular,  we consider the medium constituted of the 
lightest and most abundant pseudoscalar and vector mesons, i.e. the pions 
and $\rho$ mesons. The reactions involving the $X(3872)$ 
have already been studied in previous works, and for the sake of conciseness we 
will not reproduce them here. We refer the reader to 
Refs.~\cite{Houng:2013,Torres:2014,abreu16,QCDSR:21} for a detailed discussion.

\begin{figure}[!htbp]
	\centering
\begin{tikzpicture}
\begin{feynman}
\vertex (a1) {$\psi (p_{1})$};
	\vertex[right=1.5cm of a1] (a2);
	\vertex[right=1.cm of a2] (a3) {$\bar{D} (p_{3})$};
	\vertex[right=1.4cm of a3] (a4) {$\psi (p_{1})$};
	\vertex[right=1.5cm of a4] (a5);
	\vertex[right=1.cm of a5] (a6) {$\bar{D} (p_{3})$};
\vertex[below=1.5cm of a1] (c1) {$\pi (p_{2})$};
\vertex[below=1.5cm of a2] (c2);
\vertex[below=1.5cm of a3] (c3) {$D (p_{4})$};
\vertex[below=1.5cm of a4] (c4) {$\pi (p_{2})$};
\vertex[below=1.5cm of a5] (c5);
\vertex[below=1.5cm of a6] (c6) {$D (p_{4})$};
	\vertex[below=2cm of a2] (d2) {(a)};
	\vertex[below=2cm of a5] (d5) {(b)};
\diagram* {
  (a1) -- (a2), (a2) -- (a3), (c1) -- (c2), (c2) -- (c3), (a2) --
  [fermion, edge label'= $D^{*}$] (c2), (a4) -- (a5), (a5) -- (c6),
  (c4) -- (c5), (c5) -- (a6), (a5) -- [fermion, edge label'= $D^{*}$] (c5)
}; 
\end{feynman}
\end{tikzpicture}

\begin{tikzpicture}
\begin{feynman}
\vertex (a1) {$\psi (p_{1})$};
	\vertex[right=1.5cm of a1] (a2);
	\vertex[right=1.cm of a2] (a3) {$\bar{D} (p_{3})$};
	\vertex[right=1.4cm of a3] (a4) {$\psi (p_{1})$};
	\vertex[right=1.5cm of a4] (a5);
	\vertex[right=1.cm of a5] (a6) {$\bar{D} (p_{3})$};
\vertex[below=1.5cm of a1] (c1) {$\pi (p_{2})$};
\vertex[below=1.5cm of a2] (c2);
\vertex[below=1.5cm of a3] (c3) {$D^{*} (p_{4})$};
\vertex[below=1.5cm of a4] (c4) {$\pi (p_{2})$};
\vertex[below=1.5cm of a5] (c5);
\vertex[below=1.5cm of a6] (c6) {$D^{*} (p_{4})$};
	\vertex[below=2cm of a2] (d2) {(c)};
	\vertex[below=2cm of a5] (d5) {(d)};
\diagram* {
  (a1) -- (a2), (a2) -- (a3), (c1) -- (c2), (c2) -- (c3), (a2) --
  [fermion, edge label'= $D$] (c2), (a4) -- (a5), (a5) -- (c6),
  (c4) -- (c5), (c5) -- (a6), (a5) -- [fermion, edge label'= $D^{*}$] (c5)
}; 
\end{feynman}
\end{tikzpicture}

\begin{tikzpicture}
\begin{feynman}
\vertex (a1) {$\psi (p_{1})$};
	\vertex[right=1.5cm of a1] (a2);
	\vertex[right=1.cm of a2] (a3) {$\bar{D} (p_{3})$};
	\vertex[right=1.4cm of a3] (a4) {$\psi (p_{1})$};
	\vertex[right=1.5cm of a4] (a5);
	\vertex[right=1.cm of a5] (a6) {$\bar{D}^{*} (p_{3})$};
\vertex[below=1.5cm of a1] (c1) {$\pi (p_{2})$};
\vertex[below=1.5cm of a2] (c2);
\vertex[below=1.5cm of a3] (c3) {$D^{*} (p_{4})$};
\vertex[below=1.5cm of a4] (c4) {$\pi (p_{2})$};
\vertex[below=1.5cm of a5] (c5);
\vertex[below=1.5cm of a6] (c6) {$D^{*} (p_{4})$};
	\vertex[below=2cm of a2] (d2) {(e)};
	\vertex[below=2cm of a5] (d5) {(f)};
\diagram* {
  (a1) -- (a2), (a2) -- (a3), (c1) -- (c2), (c2) -- (c3), (a2) --
  [fermion, edge label'= $D^{*}$] (c2), (a4) -- (a5), (a5) -- (a6),
  (c4) -- (c5), (c5) -- (c6), (a5) -- [fermion, edge label'= $D$] (c5)
}; 
\end{feynman}
\end{tikzpicture}

\begin{tikzpicture}
\begin{feynman}
\vertex (a1) {$\psi (p_{1})$};
	\vertex[right=1.5cm of a1] (a2);
	\vertex[right=1.cm of a2] (a3) {$\bar{D}^{*} (p_{3})$};
	\vertex[right=1.4cm of a3] (a4) {$\psi (p_{1})$};
	\vertex[right=1.5cm of a4] (a5);
	\vertex[right=1.cm of a5] (a6) {$\bar{D}^{*} (p_{3})$};
\vertex[below=1.5cm of a1] (c1) {$\pi (p_{2})$};
\vertex[below=1.5cm of a2] (c2);
\vertex[below=1.5cm of a3] (c3) {$D^{*} (p_{4})$};
\vertex[below=1.5cm of a4] (c4) {$\pi (p_{2})$};
\vertex[below=1.5cm of a5] (c5);
\vertex[below=1.5cm of a6] (c6) {$D^{*} (p_{4})$};
	\vertex[below=2cm of a2] (d2) {(g)};
	\vertex[below=2cm of a5] (d5) {(h)};
\diagram* {
  (a1) -- (a2), (a2) -- (c3), (c1) -- (c2), (c2) -- (a3), (a2) --
  [fermion, edge label'= $D$] (c2), (a4) -- (a5), (a5) -- (a6),
  (c4) -- (c5), (c5) -- (c6), (a5) -- [fermion, edge label'= $D^{*}$] (c5)
}; 
\end{feynman}
\end{tikzpicture}

\begin{tikzpicture}
\begin{feynman}
\vertex (a1) {$\psi (p_{1})$};
	\vertex[right=1.5cm of a1] (a2);
	\vertex[right=1.cm of a2] (a3) {$\bar{D}^{*} (p_{3})$};
\vertex[below=1.5cm of a1] (c1) {$\pi (p_{2})$};
\vertex[below=1.5cm of a2] (c2);
\vertex[below=1.5cm of a3] (c3) {$D^{*} (p_{4})$};
	\vertex[below=2cm of a2] (d2) {(i)};
\diagram* {
  (a1) -- (a2), (a2) -- (c3), (c1) -- (c2), (c2) -- (a3), (a2) --
  [fermion, edge label'= $D^{*}$] (c2)
}; 
\end{feynman}
\end{tikzpicture}
	\caption{Born diagrams for the processes   
$\psi(2S) \pi \rightarrow \bar{D} D$  [(a) and (b)],
$\psi(2S) \pi \rightarrow \bar{D}^{*} D$  [(c) - (e)],
and $\psi(2S) \pi \rightarrow \bar{D}^{*} D^{*}$  [(f) - (i)]. Here t
The particle charges are not specified. We denote $ \psi(2S) \equiv \psi $. }
\label{DIAG1}
\end{figure}

\begin{figure}[!htbp]
	\centering
\begin{tikzpicture}
\begin{feynman}
\vertex (a1) {$\psi (p_{1})$};
	\vertex[right=1.5cm of a1] (a2);
	\vertex[right=1.cm of a2] (a3) {$\bar{D} (p_{3})$};
	\vertex[right=1.4cm of a3] (a4) {$\psi (p_{1})$};
	\vertex[right=1.5cm of a4] (a5);
	\vertex[right=1.cm of a5] (a6) {$\bar{D} (p_{3})$};
\vertex[below=1.5cm of a1] (c1) {$\rho (p_{2})$};
\vertex[below=1.5cm of a2] (c2);
\vertex[below=1.5cm of a3] (c3) {$D (p_{4})$};
\vertex[below=1.5cm of a4] (c4) {$\rho (p_{2})$};
\vertex[below=1.5cm of a5] (c5);
\vertex[below=1.5cm of a6] (c6) {$D (p_{4})$};
	\vertex[below=2cm of a2] (d2) {(j)};
	\vertex[below=2cm of a5] (d5) {(k)};
\diagram* {
  (a1) -- (a2), (a2) -- (a3), (c1) -- (c2), (c2) -- (c3), (a2) --
  [fermion, edge label'= $D$] (c2), (a4) -- (a5), (a5) -- (c6),
  (c4) -- (c5), (c5) -- (a6), (a5) -- [fermion, edge label'= $D$] (c5)
}; 
\end{feynman}
\end{tikzpicture}

\begin{tikzpicture}
\begin{feynman}
\vertex (a1) {$\psi (p_{1})$};
	\vertex[right=1.5cm of a1] (a2);
	\vertex[right=1.cm of a2] (a3) {$\bar{D} (p_{3})$};
	\vertex[right=1.4cm of a3] (a4) {$\psi (p_{1})$};
	\vertex[right=1.5cm of a4] (a5);
	\vertex[right=1.cm of a5] (a6) {$\bar{D} (p_{3})$};
\vertex[below=1.5cm of a1] (c1) {$\rho (p_{2})$};
\vertex[below=1.5cm of a2] (c2);
\vertex[below=1.5cm of a3] (c3) {$D (p_{4})$};
\vertex[below=1.5cm of a4] (c4) {$\rho (p_{2})$};
\vertex[below=1.5cm of a5] (c5);
\vertex[below=1.5cm of a6] (c6) {$D (p_{4})$};
	\vertex[below=2cm of a2] (d2) {(l)};
	\vertex[below=2cm of a5] (d5) {(m)};
\diagram* {
  (a1) -- (a2), (a2) -- (a3), (c1) -- (c2), (c2) -- (c3), (a2) --
  [fermion, edge label'= $D^*$] (c2), (a4) -- (a5), (a5) -- (c6),
  (c4) -- (c5), (c5) -- (a6), (a5) -- [fermion, edge label'= $D^*$] (c5)
}; 
\end{feynman}
\end{tikzpicture}

\begin{tikzpicture}
\begin{feynman}
\vertex (a1) {$\psi (p_{1})$};
	\vertex[right=1.5cm of a1] (a2);
	\vertex[right=1.cm of a2] (a3) {$\bar{D} (p_{3})$};
	\vertex[right=1.4cm of a3] (a4) {$\psi (p_{1})$};
	\vertex[right=1.5cm of a4] (a5);
	\vertex[right=1.cm of a5] (a6) {$\bar{D} (p_{3})$};
\vertex[below=1.5cm of a1] (c1) {$\rho (p_{2})$};
\vertex[below=1.5cm of a2] (c2);
\vertex[below=1.5cm of a3] (c3) {$D^* (p_{4})$};
\vertex[below=1.5cm of a4] (c4) {$\rho (p_{2})$};
\vertex[below=1.5cm of a5] (c5);
\vertex[below=1.5cm of a6] (c6) {$D^* (p_{4})$};
	\vertex[below=2cm of a2] (d2) {(n)};
	\vertex[below=2cm of a5] (d5) {(o)};
\diagram* {
  (a1) -- (a2), (a2) -- (a3), (c1) -- (c2), (c2) -- (c3), (a2) --
  [fermion, edge label'= $D$] (c2), (a4) -- (a5), (a5) -- (c6),
  (c4) -- (c5), (c5) -- (a6), (a5) -- [fermion, edge label'= $D$] (c5)
}; 
\end{feynman}
\end{tikzpicture}

\begin{tikzpicture}
\begin{feynman}
\vertex (a1) {$\psi (p_{1})$};
	\vertex[right=1.5cm of a1] (a2);
	\vertex[right=1.cm of a2] (a3) {$\bar{D} (p_{3})$};
	\vertex[right=1.4cm of a3] (a4) {$\psi (p_{1})$};
	\vertex[right=1.5cm of a4] (a5);
	\vertex[right=1.cm of a5] (a6) {$\bar{D} (p_{3})$};
\vertex[below=1.5cm of a1] (c1) {$\rho (p_{2})$};
\vertex[below=1.5cm of a2] (c2);
\vertex[below=1.5cm of a3] (c3) {$D^* (p_{4})$};
\vertex[below=1.5cm of a4] (c4) {$\rho (p_{2})$};
\vertex[below=1.5cm of a5] (c5);
\vertex[below=1.5cm of a6] (c6) {$D^* (p_{4})$};
	\vertex[below=2cm of a2] (d2) {(p)};
	\vertex[below=2cm of a5] (d5) {(q)};
\diagram* {
  (a1) -- (a2), (a2) -- (a3), (c1) -- (c2), (c2) -- (c3), (a2) --
  [fermion, edge label'= $D^*$] (c2), (a4) -- (a5), (a5) -- (c6),
  (c4) -- (c5), (c5) -- (a6), (a5) -- [fermion, edge label'= $D^*$] (c5)
}; 
\end{feynman}
\end{tikzpicture}

\begin{tikzpicture}
\begin{feynman}
\vertex (a1) {$\psi (p_{1})$};
	\vertex[right=1.5cm of a1] (a2);
	\vertex[right=1.cm of a2] (a3) {$\bar{D}^* (p_{3})$};
	\vertex[right=1.4cm of a3] (a4) {$\psi (p_{1})$};
	\vertex[right=1.5cm of a4] (a5);
	\vertex[right=1.cm of a5] (a6) {$\bar{D}^* (p_{3})$};
\vertex[below=1.5cm of a1] (c1) {$\rho (p_{2})$};
\vertex[below=1.5cm of a2] (c2);
\vertex[below=1.5cm of a3] (c3) {$D^* (p_{4})$};
\vertex[below=1.5cm of a4] (c4) {$\rho (p_{2})$};
\vertex[below=1.5cm of a5] (c5);
\vertex[below=1.5cm of a6] (c6) {$D^* (p_{4})$};
	\vertex[below=2cm of a2] (d2) {(r)};
	\vertex[below=2cm of a5] (d5) {(s)};
\diagram* {
  (a1) -- (a2), (a2) -- (a3), (c1) -- (c2), (c2) -- (c3), (a2) --
  [fermion, edge label'= $D$] (c2), (a4) -- (a5), (a5) -- (c6),
  (c4) -- (c5), (c5) -- (a6), (a5) -- [fermion, edge label'= $D$] (c5)
}; 
\end{feynman}
\end{tikzpicture}

\begin{tikzpicture}
\begin{feynman}
\vertex (a1) {$\psi (p_{1})$};
	\vertex[right=1.5cm of a1] (a2);
	\vertex[right=1.cm of a2] (a3) {$\bar{D}^* (p_{3})$};
	\vertex[right=1.4cm of a3] (a4) {$\psi (p_{1})$};
	\vertex[right=1.5cm of a4] (a5);
	\vertex[right=1.cm of a5] (a6) {$\bar{D}^* (p_{3})$};
\vertex[below=1.5cm of a1] (c1) {$\rho (p_{2})$};
\vertex[below=1.5cm of a2] (c2);
\vertex[below=1.5cm of a3] (c3) {$D^* (p_{4})$};
\vertex[below=1.5cm of a4] (c4) {$\rho (p_{2})$};
\vertex[below=1.5cm of a5] (c5);
\vertex[below=1.5cm of a6] (c6) {$D^* (p_{4})$};
	\vertex[below=2cm of a2] (d2) {(t)};
	\vertex[below=2cm of a5] (d5) {(u)};
\diagram* {
  (a1) -- (a2), (a2) -- (a3), (c1) -- (c2), (c2) -- (c3), (a2) --
  [fermion, edge label'= $D^*$] (c2), (a4) -- (a5), (a5) -- (c6),
  (c4) -- (c5), (c5) -- (a6), (a5) -- [fermion, edge label'= $D^*$] (c5)
}; 
\end{feynman}
\end{tikzpicture}

	\caption{Born diagrams for the processes   
$\psi(2S) \rho \rightarrow \bar{D} D$  [(j) - (m)],
$\psi(2S) \rho \rightarrow \bar{D}^{*} D$  [(n) - (q)] and
$\psi(2S) \rho \rightarrow \bar{D}^{*} D^{*}$  [(r) - (u)].  The particle charges are not specified. We denote $ \psi(2S) \equiv \psi $.}
\label{DIAG2}
\end{figure}
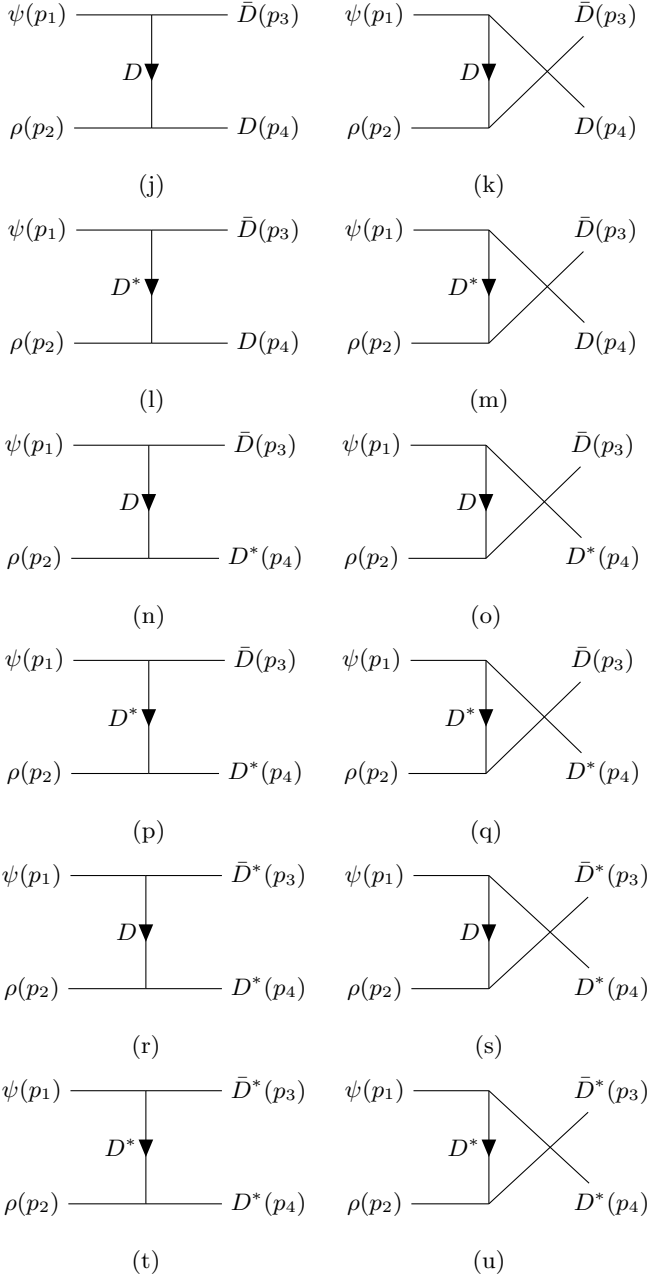

To the best of our knowledge, in contrast to the case of the $J/\psi$    
~\cite{Yongseok: 2001, Navarra:2005, Abreu:2018}, 
there is no effective theory for the reactions involving the $\psi(2S)$. 
In the lack of works on this subject, we adopt an effective 
approach, based on the $J/\psi$ studies, to describe the reactions 
$\psi(2S) \pi \rightarrow \bar{D}^{(*)} D^{(*)}$ and 
$\psi(2S) \rho \rightarrow \bar{D}^{(*)} D^{(*)}$, as well as the inverse 
processes. In  Figs.~\ref{DIAG1} and~\ref{DIAG2} we show the lowest-order
Born diagrams contributing to each process.
To calculate their respective cross sections, we will adapt the  effective Lagrangians introduced in~\cite{ Yongseok: 2001, Navarra:2005, Abreu:2018}. The couplings involving the $\psi (2S)$ and the $ D^{\ast} $ mesons are (we denote $ \psi(2S) \equiv \psi $), 
\begin{eqnarray}
	\mathcal{L}_{\psi  D D} & = & g_{\psi DD} \psi_{\mu} ( \partial^{\mu} D \bar{D} – D \partial^{\mu} \bar{D} ),
\nonumber \\
	\nonumber \mathcal{L}_{\psi D^{*} D^{*}} &=& -i g_{\psi D^{*} D^{*}}  [ \psi^{\mu} ( \partial_{\mu} D^{*\nu} \bar{D}_{\nu}^{*} – D^{*\nu} \partial_{\mu} \bar{D}_{\nu}^{*})
	\nonumber \\
	\nonumber  & &+(\partial_{\mu} \psi_{\nu} D^{*\nu} - \psi_{\mu} \partial_{\nu} D^{*\nu} )\bar{D}^{*\mu}
	\nonumber \\
	& &+D^{*\mu} ( \psi^{\nu} \partial_{\mu} \bar{D}_{\nu}^{*}  - \partial_{\mu} \psi_{\nu} \bar{D}^{*\nu} ) ],
	\nonumber \\
	\mathcal{L}_{\psi D D^{*}} & =  & - g_{\psi D D^{*}} \epsilon^{\mu \nu \alpha \beta} \partial_{\mu} \psi_{\nu} ( \partial_{\alpha} D_{\beta}^{*}  \bar{D} + D \partial_{\alpha} \bar{D}_{\beta}^{*} ); \nonumber \\
\label{EffLagf1}
\end{eqnarray}
while the vertices involving the $ D^{\ast} $ mesons and the pions and $\rho$ mesons are 
\begin{eqnarray} 
\mathcal{L}_{\pi D D^*} & = & ig_{\pi D D^*} D^{* \mu} \vec{\tau} \cdot (\bar{D} \partial_\mu \vec{\pi} - \partial_{\mu} \bar{D} \vec{\pi}) + h.c., \nonumber \\ 
\mathcal{L}_{\rho DD} & = & ig_{\rho DD} (D \vec{\tau } \partial_{\mu} \bar{D} - \partial_{\mu} D \vec{\tau} \bar{D}) \cdot \vec{\rho}^{\mu}, \nonumber \\
\mathcal{L}_{\rho D^* D^*} &= & i g_{\rho D^* D^*} \left[ (\partial_{\mu} D^{* \nu} \vec{\tau } \bar{D}^{*}_{\nu} - D^{* \nu} \vec{\tau } \partial_{\mu} \bar{D}^{*}_{\nu}) \cdot \vec{\rho}^{\mu} \right. \nonumber \\
&  & + (D^{* \nu} \vec{\tau } \cdot \partial_{\mu} \vec{\rho}_{\nu} - \partial_{\mu} D^{* \nu} \vec{\tau } \cdot \vec{\rho}_{\nu}) \bar{D}^{* \mu}\nonumber \\
  & & + \left. D^{*\mu} (\vec{\tau } \cdot \vec{\rho}^{\nu} \partial_{\mu} \bar{D}^{*}_{\nu} - \vec{\tau } \cdot \partial_{\mu} \vec{\rho}^{\nu} \bar{D}^{*}_{\nu})\right] ,\nonumber\\ 
\mathcal{L}_{\pi D^{*} D^{*}} &=& -g_{\pi D^* D^*} \epsilon^{\mu \nu \alpha \beta} \partial_{\mu} D^{8}_{\nu} \pi \partial_{\alpha} \bar{D}^{*}_{\beta},\nonumber\\ 
\mathcal{L}_{\rho DD^*} &=& -g_{\rho DD^*} \epsilon^{\mu \nu \alpha \beta} (D \partial_{\mu} \rho_{\nu} \partial_{\alpha} \bar{D}^{*}_{\beta} + \partial_{\mu} D^{*}_{\nu} \partial_{\alpha} \rho_{\beta} \bar{D}).
	\nonumber \\
\label{EffLagf2}
\end{eqnarray}
In Eqs.~(\ref{EffLagf1}) and~(\ref{EffLagf2}), $D^{(\ast)}$ and $\bar{D}^{(*)} $  represent the isospin doublets for the pseudoscalar (vector) charmed mesons; 
$ \psi_{\nu} $ denotes the  $\psi (2S)$ vector field; $\vec{\tau}$ represents the Pauli matrices in the isospin space; $\vec{\pi}$ and $\vec{\rho}$      
are the pion and $ \rho $ isospin triplets, respectively. The coupling constants $g_{\pi DD^*}$, $g_{\rho DD}$,
$g_{\rho D^* D^*}$, $g_{\pi D^* D^*}$ and $g_{\rho D D^*}$ will be discussed below.

These effective Lagrangians introduced above allow us to determine the amplitudes
of the $\psi(2S)$ absorption processes depicted in Figs.~\ref{DIAG1}
and~\ref{DIAG2},
\begin{eqnarray}
	\mathcal{M}_{\psi \pi \rightarrow \bar{D} D} &\equiv & \sum_{{i} = a,b} \mathcal{M}^{(  \textbf{i} )},
	\nonumber \\
	\mathcal{M}_{\psi \pi \rightarrow \bar{D} D^{*}} &\equiv & \sum_{\textbf{i} = c,d,e} \mathcal{M}^{(\textbf{i})},
	\nonumber \\
	\mathcal{M}_{\psi \pi \rightarrow \bar{D}^{*} D^{*}} &\equiv & \sum_{\textbf{i}  = f,g,h,i} \mathcal{M}^{( \textbf{i} )},
	\nonumber \\
	\mathcal{M}_{\psi \rho \rightarrow \bar{D} D} &\equiv & \sum_{\textbf{i} = j,k,l,m} \mathcal{M}^{(  \textbf{i} )},
	\nonumber \\
	\mathcal{M}_{\psi \rho \rightarrow \bar{D} D^{*}} &\equiv & \sum_{\textbf{i} = n,o,p,q} \mathcal{M}^{(  \textbf{i} )},
	\nonumber \\
	\mathcal{M}_{\psi \rho \rightarrow \bar{D}^{*} D^{*}} &\equiv & \sum_{\textbf{i} = r,s,t,u} \mathcal{M}^{(  \textbf{i} )},
\label{eq:Amplitudes}
\end{eqnarray}
where $\mathcal{M}^{ (\textbf{i} ) }$ denotes the contribution coming from the specific reaction $(i)$; the expressions are explicitly summarized in Appendix~\ref{Ampl}.

As usual, we employ form factors to prevent the artificial growth of the 
amplitudes with energy and to take into account the finite size of hadrons.  To the best of our knowledge there is so far no systematic study of the couplings of Eqs.~(\ref{EffLagf1}) and~(\ref{EffLagf2}) involving the $\psi(2S)$. Due to the similarities of these vertices with those involving the $J/\psi$, we assume that the form factors for 
the $\psi(2S)$ are given by the same parametrizations of those for the $J/\psi$, which have already been computed with QCD 
sum rules in Refs.~\cite{QCDSR:23,QCDSR:18, QCDSR:19, QCDSR:20, QCDSR:21, Navarra:2005}. They are given by  
\begin{eqnarray}
I) \,\,\,\, g_{M_1 M_2 M_3} & = & \frac{A}{Q^2 + B},
\label{eq:FF1} \nonumber \\ 
II) \,\,\,\, g_{M_1 M_2 M_3} & = & A \, e^{ - \left( \frac{ B + Q^2 }{C} \right)  } , 
\label{eq:FF2}
\end{eqnarray}
where the values of the constants appearing above  are given in Table \ref{Table:Parameters-QCDSR}. The cross sections should depend on 
the size of the interacting hadron.  From the studies of charmonium spectroscopy, it is well known that the $\psi(2S)$ is larger than the fundamental charmonium state by a factor of about 2 (see for example the discussion in~\cite{Godfrey:1985xj}). Therefore, based on geometrical arguments we expect the $\psi(2S)$ to have bigger cross sections than the $J/\psi$ by a factor of about 4. Hence, in the lack of reliable  estimates of the coupling constants and form factors for the $\psi(2S)$ vertices, we use the parametrizations given in Eq.~(\ref{eq:FF2}) 
and Table \ref{Table:Parameters-QCDSR}  but with the coupling constants varying in the range $[g_{M_1 M_2 M_3}, 2g_{M_1 M_2 M_3}]$. This 
will be the main source of uncertainties in our calculation.

\begin{table}[h!] 
\begin{center} 
	\caption{Form factors $g_{M_1 M_2 M_3}$
for the respective vertices $M_1 M_2 M_3$ presented in Eq. (\ref{eq:FF2}), computed 
with  QCD sum rules in Refs.
~\cite{QCDSR:23,QCDSR:18, QCDSR:19, QCDSR:20, QCDSR:21, Navarra:2005} for the $J/\psi$. $M_2$ denotes the exchange particle.}
\begin{tabular}{ccccc} 
	\hline 
	\hline 
$M_1$ $M_2$ $M_3$ & Form & A & B & C \\ 
	\hline 
$\psi D D$      & II & 5.8 & 20 & 15.8  \\
$\psi D^* D$      & II & 20 & 27 & 18.6  \\
$\psi D D^* $      & II & 13 & 26 & 21.2  \\ 
$\psi D^* D^* $      & II & 6.2 & 0 & 3.55  \\ 
$\pi D D^* $      & I    & 126  & 11.9 & -  \\ 
$\pi D^* D $      & I    & 126  & 11.9 & -  \\ 
$\rho D D$        & I    & 37.5 & 12.1 & - \\ 
$\rho D^*  D^*$    & II   & 4.9  & 0 & 13.3  \\ 
$\pi D^*  D^*$     & II   & 4.8  & 0 & 6.8   \\ 
$\rho D D^*$      & I    & 234  & 44 & -    \\ 
$\rho D^* D$      & I    & 234  & 44 & -    \\ 
	\hline 
\end{tabular} 
\label{Table:Parameters-QCDSR} 
\end{center} 
\end{table}

\section{Cross sections for the processes involving $\psi (2S)$}
\label{CrossSections} 

\subsection{Vaccum cross sections}

The isospin-spin-averaged cross section in the center-of-mass (CM) frame for a specific $ ab \rightarrow cd $ process is given by
\begin{eqnarray}
\sigma_{ab \rightarrow cd} = \frac{1}{64 \pi^2 s g_{a} g_{b}} \frac{ | \vec{p}_{cd} |}{\vec{p}_{ab} |}  \int d  \Omega \sum_{S, I} | \mathcal{M}_{ab \rightarrow cd} (s,\theta) |^2  
\label{eq:CrossSection}
\end{eqnarray}
where $\sqrt{s}$ is the CM energy; $|\vec{p}_{ab}|$ and $|\vec{p}_{cd}|$ 
are the magnitudes of three-momenta of initial and final particles in the CM frame, 
respectively; $\sum_{S,I} $ means the sum over the
spins and isospins of the particles in the initial and 
final state, weighted by the isospin and spin degeneracy factors $g_a = (2S_a + 1)(2I_a +  1)$ and  $g_b = (2S_b + 1)(2I_b +  1)$ of the two
particles forming the initial state. The  cross sections for the respective inverse processes can be computed through the use of the detailed balance relation,
\begin{eqnarray}
\sigma_{cd \rightarrow ab} (s) = \frac{g_a g_b }{g_c g_d } \frac{ |\vec{p}_{ab} |}{ |\vec{p}_{cd} |} \sigma_{ab  \rightarrow cd} (s).
\label{eq:CrossSectionInverse}
\end{eqnarray}

The calculations are performed using the isospin-averaged masses of the light and  
heavy mesons according to the values reported in the PDG~\cite{Workman:2022ynf}. 
Since we use a range of values for the  $\psi (2S)$ couplings 
(due to the uncertainties) the results are shown as bands.

\begin{figure}[!htbp]
	\centering
\includegraphics[{width=8.0cm}]{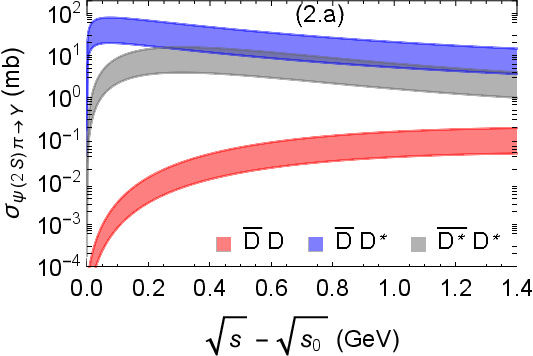} \\ 
\includegraphics[{width=8.0cm}]{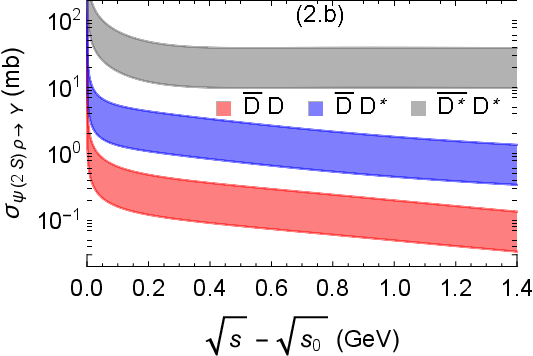} 
	\caption{ Cross sections for the absorption processes 
$\psi \pi \rightarrow \bar{D}^{(*)} D^{(*)}$
(2.a) and 
$\psi \rho \rightarrow \bar{D}^{(*)} D^{(*)}$
(2.b), as functions of the relative CM energy 
$\sqrt{s} - \sqrt{s_{0}}$. $ \sqrt{s _{0}}$ is the threshold energy of the respective channel. }
\label{Fig:csabsorption}
\end{figure}

\begin{figure}[!htbp]
	\centering
\includegraphics[{width=8.0cm}]{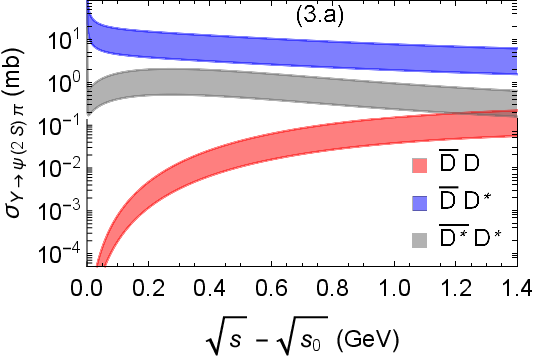} \\ 
\includegraphics[{width=8.0cm}]{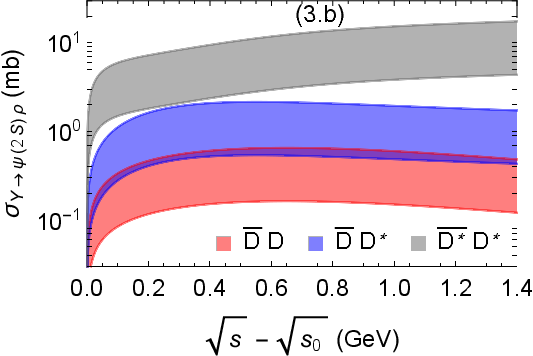} 
	\caption{ Cross sections for the production processes 
$\bar{D}^{(*)} D^{(*)} \rightarrow \psi \pi$
(3.a) and $\bar{D}^{(*)} D^{(*)} \rightarrow \psi \rho$
(3.b), as functions of the relative CM energy 
$\sqrt{s} - \sqrt{s _{0}}$.}
\label{Fig:csproduction}
\end{figure}

The cross sections for the $\psi (2S)$ absorption by comoving 
light mesons as functions of the relative CM energy
$\sqrt{s} - \sqrt{s _{0}}$ ($ \sqrt{s _{0}}$ being the threshold energy of the 
respective channel) are plotted in Fig.~\ref{Fig:csabsorption}.   
Except for the channels $\psi \pi \rightarrow \bar{D}^{*} D $ and  
$\psi \pi \rightarrow \bar{D}^{*} D^{*}$, all the cross sections are    
exothermic, showing a substantial decrease right beyond the threshold. We remark 
that the $\psi \pi \rightarrow \bar{D} D $ channel  
acquires large values very near the threshold, not visible in the plot, and   
after that suffers a strong decrease.  In the region close  to the threshold  
up to moderate energies, i.e. $\sqrt{s} - \sqrt{s _{0}} \simeq 500 \MeV$, the 
different channels present magnitudes of
the order of $  10^{-4} -  10^{1} \mb$, but with the reactions with final state  $ \bar{D} D $ being suppressed with respect to the other ones. 
In the case of the inverse processes, the cross sections are plotted in Fig.~\ref{Fig:csproduction}. As expected, very near the threshold they have an opposite behavior with respect to the corresponding $\psi (2S)$ absorption reactions in Fig.~\ref{Fig:csabsorption}. From the region close to the threshold up to moderate energies,  we observe  that the cross sections are of the order $  10^{-4} -  10^{0} \mb$, and in general those for $\psi (2S)$ absorption reactions have smaller magnitudes than their respective inverse processes. This is an important feature (since it appears in the  energy range relevant for heavy ion collisions) which can be attributed to the differences in the phase space and the degeneracies encoded in Eq.~(\ref{eq:CrossSectionInverse})~\cite{Abreu:2022jmi}.

\subsection{Thermally-averaged cross sections}

In a hadron gas the temperature determines the order of magnitude of the collision energy. Hence it is
more realistic to use the vacuum cross sections weighted by the thermal momentum distributions of the colliding particles. 
For processes with
a two-particle initial state going into two final state particles
$ab \rightarrow cd$, it is given by ~\cite{Abreu:2022EPJC, QCDSR:21, Abreu:2022jmi, Houng:2013, Koch:1986} 
\begin{eqnarray}
\langle \sigma_{a b \rightarrow c d } \upsilon_{a b}\rangle &  = & 
\frac{ \int  d^{3} \mathbf{p}_a  d^{3}
\mathbf{p}_b f_a(\mathbf{p}_a) f_b(\mathbf{p}_b) \sigma_{a b \rightarrow c d } 
\,\,\upsilon_{a b} }{ \int d^{3} \mathbf{p}_a  
d^{3} \mathbf{p}_b f_a(\mathbf{p}_a) f_b(\mathbf{p}_b) }
\nonumber \\
& = & \frac{1}{4 \beta_a ^2 K_{2}(\beta_a) \beta_b ^2 K_{2}(\beta_b) } 
\nonumber \\
& & \times \int _{z_0} ^{\infty } dz  K_{1}(z) \,\,\sigma_{ab \rightarrow cd}  (s=z^2 T^2) 
\nonumber \\
& & \times \left[ z^2 - (\beta_a + \beta_b)^2 \right]
\left[ z^2 - (\beta_a - \beta_b)^2 \right],
\nonumber \\
  \label{Eq:AvCrossSection}
\end{eqnarray}
where $\upsilon_{a b}$ denotes the relative velocity of the two initial interacting    
particles; the function $f_i(\mathbf{p}_i)$ is the Bose-Einstein distribution;
$\beta _i = m_i / T$, with $T$ being the temperature; $z_0 = max(\beta_a + \beta_b,\beta_c + \beta_d)$, and $K_1$ and $K_2$ are the modified Bessel functions of
second kind. It can be seen from the expression in the second line of Eq.~(\ref{Eq:AvCrossSection}) that the thermal average suppresses the configurations very close to the thresholds: for sufficiently large $z_0$ the $K_1(z)$ function acquires very small values.

\begin{figure}[!htbp]
	\centering
\includegraphics[{width=8.0cm}]{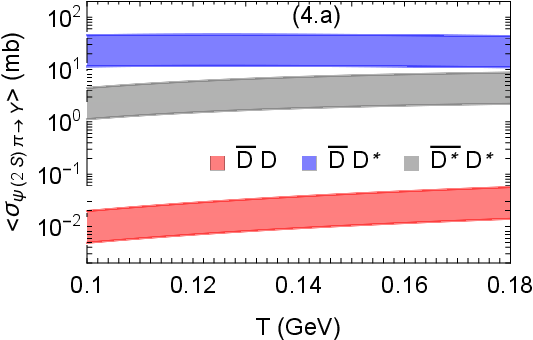} \\
\includegraphics[{width=8.0cm}]{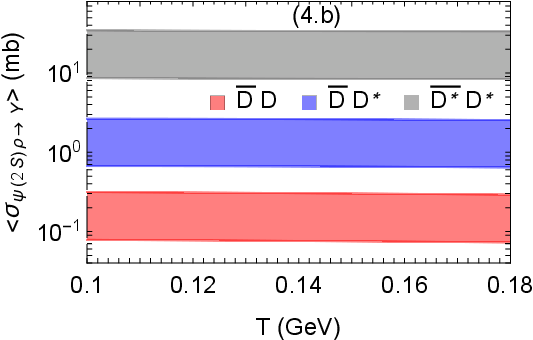} 
	\caption{ Thermally averaged cross sections for the absorption processes 
$\psi \pi \rightarrow \bar{D}^{(*)} D^{(*)}$ (4.a) and $\psi \rho \rightarrow \bar{D}^{(*)} D^{(*)}$ (4.b), as functions of the temperature.}
\label{Fig: AvCrSec-Abs}
\end{figure}

\begin{figure}[!htbp]
	\centering
\includegraphics[{width=8.0cm}]{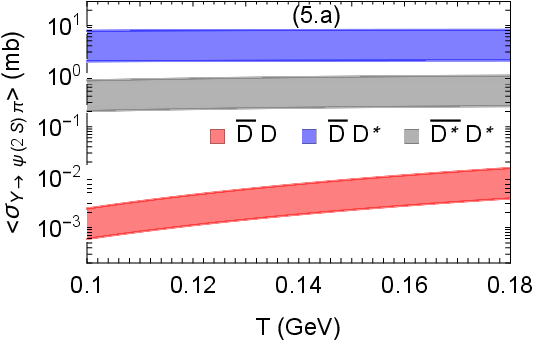} \\
\includegraphics[{width=8.0cm}]{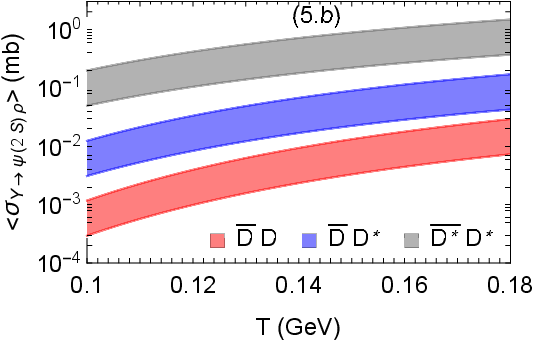} 
	\caption{ Thermally averaged cross sections for the production processes  $\bar{D}^{(*)} D^{(*)} \rightarrow \psi \pi$ (5.a) and  $\bar{D}^{(*)} D^{(*)} \rightarrow \psi \rho$ (5.b), as functions of the temperature. }
\label{Fig:AvCrSec-Prod}
\end{figure}

In Fig.~\ref{Fig: AvCrSec-Abs} we plot the thermally
averaged cross sections as functions of the temperature for the 
$\psi (2S)$ absorption by comoving light mesons using the vacuum cross-section results obtained previously. In general, they have a weak dependence on the temperature. The tendency seen in the vacuum at moderate energies is reproduced here: reactions with $ \bar{D} D $ final state are suppressed with respect to the other ones. 
On the other hand, for the  $\psi (2S)$ production processes, shown in Fig \ref{Fig:AvCrSec-Prod}, the reactions involving the production of  $\psi (2S) \rho$ present a strong dependence on the temperature. Most importantly, the absorption cross sections are always larger than  production  ones and 
the difference can reach  two orders of magnitude, depending on the temperature. According to previous studies, this feature may have strong  
impact on the final yield of the  $\psi (2S)$ in heavy ion collisions.  This issue will be addressed in the next section.

\section{Multiplicities of $\psi(2S)$ and   $X(3872)$}
\label{abundance}

\subsection{Time Evolution} \label{sec:time_ev}

Here we will present the formalism used to determine the yields of 
$\psi (2S)$ and $X(3872)$ in a hadron gas formed in the final stage of heavy ion collisions. We will use the thermally-averaged cross sections 
estimated in the previous section  for $\psi(2S)$. For $X(3872)$ we will use the results reported in Ref.~\cite{QCDSR:21} and include the contributions of the anomalous vertex $XD^{*}\bar{D}^{*}$ presented in Refs.~\cite{abreu16,Torres:2014}. These quantities 
will be employed as entries into the time-evolution equations to estimate the gain and loss terms. Explicitly, the rate equation is ~\cite{QCDSR:21, abreu16, Nielsen:2011,  EXHIC:2017, Abreu:2018, Houng:2013,Abreu:2023awj}
\begin{eqnarray} 
	\nonumber  \frac{ d N_{h}  (\tau)}{d \tau} &=& \sum_{\substack{\bar{c} = \bar{D},\bar{D}^{*}; \\ c = D, D^{*}; \\ 
\varphi = \pi, \rho}} 
\left[ \langle \sigma_{\bar{c}  c \rightarrow \varphi h} \upsilon_{\bar{c}  c} \rangle n_{\bar{c}  } (\tau) N_{c} (\tau)
 \right. 
	\nonumber \\ 
	&& -  \left.  \langle \sigma_{ \varphi h\rightarrow \bar{c}  c} \upsilon_{\psi \varphi } \rangle n_{\varphi} (\tau) N_{h }(\tau) 
\right], 
\label{Eq:RatioEquation}
\end{eqnarray}
where $N_h (\tau )$ represents the multiplicity of the state of type $h$ $(h = \psi (2S), X(3872) )$; $n_i (\tau )$ and $N_i (\tau )$ denote the density and the number of the meson of type $i$ at a given $\tau $. The pions, $\rho$ and charmed mesons in the reactions discussed previously are assumed to be in equilibrium, implying that with the Maxwell-Boltzmann approximation $n_{i} (\tau)$ becomes
\begin{eqnarray}
n_{i} (\tau) &\approx & \frac{1}{2 \pi^2}\gamma_{i} g_{i} m_{i}^2 T(\tau)K_{2}\left(\frac{m_{i} }{T(\tau)}\right), 
\label{Eq:Densities}
\end{eqnarray}
where $\gamma _i$, $g_i$ and $m_i$ is the fugacity, degeneracy factor and particle mass $i$, respectively. The multiplicity $N_i (\tau)$ is then obtained by multiplying $n_i(\tau)$ by the volume $V(\tau)$.  Following preceding works, we model the hadron gas expansion by the boost invariant Bjorken flow with an accelerated transverse expansion.  The volume and temperature profiles as a function of the proper time 
$\tau$ are as follows~\cite{abreu16, Nielsen:2011, EXHIC:2017, Abreu:2018, Houng:2013}:
\begin{eqnarray} 
V(\tau) & = & \pi \left[ R_C + \upsilon_C \left(\tau - \tau_C\right) + 
\frac{a_C}{2} \left(\tau - \tau_C \right)^2 \right]^2 \tau c ,
\nonumber \\
T(\tau) & = & T_C - \left( T_H - T_F \right) \left( \frac{\tau - 
\tau _H }{\tau _F - \tau _H}\right)^{\frac{4}{5}} ,
\label{Eq:TempVol}
\end{eqnarray}
where $R_C, \upsilon_C, a_C$ and  $T_C$ are the transverse size, transverse velocity, transverse acceleration and temperature at the time $\tau_C$, respectively; $T_H (T_F)$ is the temperature at the hadronization (kinetic freeze-out) time $\tau_H (\tau_F)$. 
The parameters in Eq. (\ref{Eq:TempVol}) are fixed according to Ref.~\cite{EXHIC:2017} for a hadronic medium formed in central Pb Pb collisions at $\sqrt{s_{NN}} = 5.02$ TeV; for completeness they are given in Table~\ref{Table:param}. 

\begin{center}
\begin{table}[!htbp]
\caption{In the first three rows we list the set of parameters used in Eq.~(\ref{Eq:TempVol}) for the hydrodynamic expansion and cooling of the hadronic medium formed in central PbPb collisions at $\sqrt{s_{NN}} = 5.02$ TeV~\cite{EXHIC:2017}. In the fourth row we list the multiplicity of the charm quark and light mesons used in the model. In the last two rows the quark masses and the frequency used in the coalescence model are listed~\cite{QCDSR:21}. }
\vskip1.5mm
\label{param}
\begin{tabular}{ c c c }
\hline
\hline
 $v_C$ (c) & $a_C$ (c$^2$/fm) & $R_C$ (fm)   \\   
0.5 & 0.09 & 11  
\\  
\hline
 $\tau_C$ (fm/c) & $\tau_H$ (fm/c)  &  $\tau_F$ (fm/c)  \\   
7.1  & 10.2 & 21.5
\\  
\hline
  $T_C (\MeV)$  & $T_H (\MeV)$ & $T_F (\MeV)$ \\   
 156 & 156 & 115   \\  
\hline
 $N_c$  & $N_{\pi}(\tau_F)$ & $N_{\rho}(\tau_H)$ \\   
 14 & 2410 & 184 \\  
 \hline
 $V_C $ (fm${}^3$)  & $m_q (\MeV)$ & $m_c  (\MeV)$ \\   
   5380  & 350 &  1500  \\  
 \hline
 $\omega (\MeV)$  &  &  \\   
  220  &  &    \\  
\hline
\hline
\end{tabular}
\label{Table:param}
\end{table}
\end{center}

Additionally, the multiplicities of the light mesons as well as of the charm quarks in charmed mesons are also shown in Table~\ref{Table:param}. For the light mesons, their fugacities in Eq.~(\ref{Eq:Densities}) appear as normalization parameters to adjust the multiplicities given in Table~\ref{Table:param}. In the case of charm quarks, since they are produced in the early stages of the collision, we assume that the total number of charm quarks $(N_c)$ in charmed hadrons remains approximately conserved during the hydrodynamic expansion, which leads to the condition $ N_c = n_c (\tau) \times V(\tau) = const.$, engendering a time-dependent charm-quark-fugacity factor $\gamma _c \equiv \gamma _c (\tau)$ in Eq.~(\ref{Eq:Densities}).

To fix the initial condition for the yield $N_{h}  (\tau)$ appearing in the integro-differential equation (\ref{Eq:RatioEquation}), i.e. the yield of the state $h$ at the end of QGP, we employ the so-called coalescence model, which is characterized by the convolution of the density matrix of the constituents of $h$ with its Wigner function~\cite{QCDSR:21,EXHIC:2017,Abreu:2021}. This model has the advantage of carrying information on the intrinsic structure of the system, such as angular momentum and the type and number of constituent quarks. More concretely, according to this approach the yield of a hadronic state of type $h$ can be written as 
\begin{eqnarray}
\nonumber N_{h} &\approx & g_{h} \prod_{ j=1 }^{ n } \frac{ N_j }{ g_j } \prod_{ i=1 }^{ n-1 } 
\frac{ (4 \pi \sigma_i^2 )^{ \frac{ 3 }{ 2 }} }{ V(\tau) (1 + 2 \mu_i T(\tau) \sigma_i^2 )} 
\nonumber \\
	& & \times \left[ \frac{ 4 \mu_i T(\tau) \sigma_i^2 }{ 3 (1 + 2 \mu_i T(\tau) \sigma_i^2 ) } \right]^{l_i}, 
\label{coalmod}
\end{eqnarray}
where $g_j$ and $N_j$ are the degeneracy and number of the $j$-th constituent of $h$ and
$\sigma_i = (\mu_i \omega)^{-1/2}$; the quantity $\omega$ is the oscillator frequency (taking an harmonic oscillator as a model of the hadron internal structure) and
$\mu$ the reduced mass, i.e. $\mu^{-1} = m_{i+1}^{-1} + \left( \sum_{j=1}^{i} m_j \right)^{-1}$; $l_i$ is the angular momentum of the system: it is 0 for an $S$ wave and 1 for a $P$ wave. Table~\ref{Table:param} summarizes the oscillation frequency for the charmed hadrons, the charm quark number and masses used here. 

From Eq.~(\ref{coalmod}) we have the following multiplicities for $\psi(2S)$ and $X(3872) $ at the end of the mixed phase: 
\begin{eqnarray}
\nonumber N_{\psi(2S)} (\tau_H) & \approx &   1.8 \times 10^{-4}, \\
\nonumber  N_{X}^{(4q)} (\tau_H) &\approx &   1.8 \times 10^{-4}, \nonumber \\ 
N_{X}^{(Mol.)} (\tau_H) &\approx &   1.1 \times 10^{-2}, 
\label{coalmod2}
\end{eqnarray}
In the case of the molecular state $X(3872)^{(Mol.)}$, to calculate the oscillation frequency, we have employed the expression $\omega = 6B$, with $B$ being the binding energy of the $X(3872)$ considered as a  $(D^0\bar D^{*0} +c.c.)$ bound state. As it can be seen, the coalescence mechanism generates initial conditions in which molecules are more abundant than compact tetraquarks by a factor of about 60, reflecting the fact that forming  a loosely bound state is easier than a compact tetraquark.

\begin{figure}[!htbp]
	\centering
\includegraphics[{width=8.0cm}]{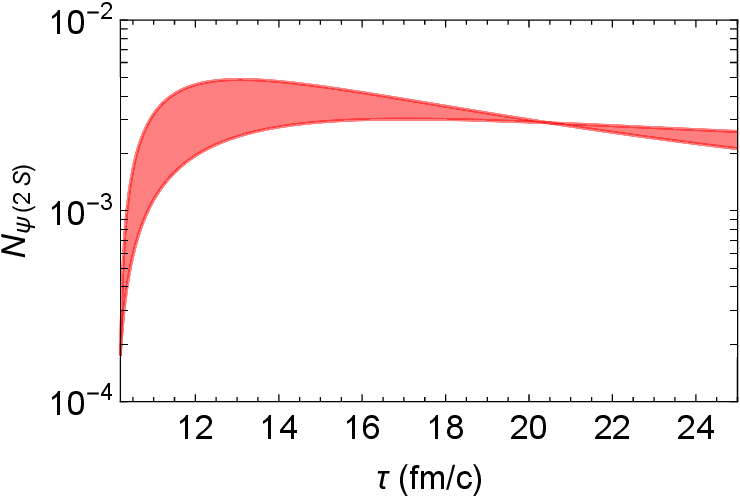} \\
\includegraphics[{width=8.0cm}]{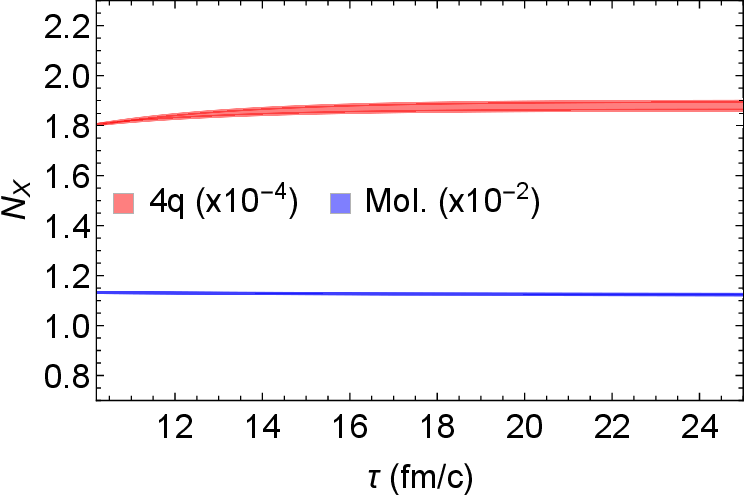} \\
\includegraphics[{width=8.0cm}]{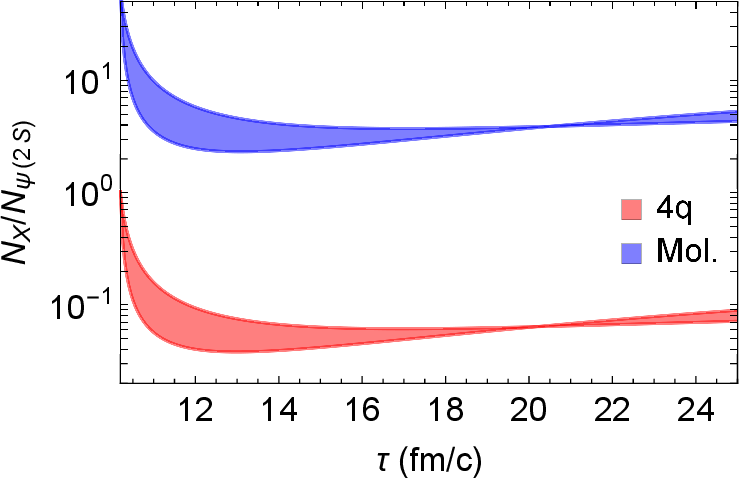} 
	\caption{Multiplicity of $\psi(2S)$ (upper panel), of $X(3872)$ 
(central panel) and their ratio (lower panel) as a function of the 
 proper time in central Pb Pb collisions at $\sqrt{s_{NN}} =$ 5.02 TeV. }
\label{Fig:Abundance-Time}
\end{figure}

In Fig. \ref{Fig:Abundance-Time} we show the time evolution of the $\psi(2S)$ and $X(3872)$ multiplicities as a function of the proper time. In the case of $\psi(2S)$, the final yield increases by a factor of about one order of magnitude, which means that the gain terms in the evolution equation (\ref{Eq:RatioEquation}) play a dominant role at higher temperatures. Strictly speaking, at the beginning of the hadron gas phase the densities and multiplicities of the charmed mesons are sufficient to counterbalance the bigger magnitudes of the thermal cross sections for the absorption processes multiplied by the densities of the light mesons and the multiplicity of $\psi(2S)$. As the time evolves, the gain and loss terms become almost of the same order (considering the uncertainties), and the $\psi (2S)$ multiplicity suffers just a slight reduction.

For the $X(3872)$ abundance, we remark that its time evolution has already been analyzed in Refs.~\cite{Houng:2013,abreu16}  for central Au Au collisions at $\sqrt{s_{NN}} = 200$ GeV. Very recently, in Ref.~\cite{QCDSR:21} this analysis has been redone for central Pb Pb collisions at $\sqrt{s_{NN}} = 5.02$ TeV, for the purpose of comparison between the final yields of $T_{cc}^+(3985)$ and $X(3872)$. Since we are interested in the $X(3872)$ to  $\psi(2S)$ yield ratio, for completeness we present again in Fig.~\ref{Fig:Abundance-Time} the $N_{X(3872)}(\tau)$. 
We improved the calculation including  the anomalous vertex $XD^{*}\bar{D}^{*}$ as discussed in~\cite{Torres:2014}. As can be seen from the figure, tetraquarks reach a final yield which is much smaller 
(by a factor of about 50) than the one of $(D \bar D^* + c.c.)$  molecules.  Another feature is that in both cases the $X(3872)$ abundance does not present a sizable change with time.

We also show in Fig.~\ref{Fig:Abundance-Time} the time evolution of the $X(3872)$ 
to $\psi(2S)$ yield ratio. As expected from the discussion above, we observe a  
strong variation of this ratio in the early times  and after that a relative     
stabilization. In Ref.~\cite{Yun:2022evm} the coalescence model has been used to 
estimate the $N_X$ and the statistical hadronization model to obtain 
$N_{\Psi(2S)}$. However, the hadronic interactions have 
not been taken into account. This has resulted in a ratio with central value of 
0.806 for the $X(3872)$ in the molecular configuration, argued to be consistent 
with the one observed by the CMS Collaboration. As shown in                
Fig.~\ref{Fig:Abundance-Time} we find a bigger ratio for the molecular 
configuration than that of Ref.~\cite{Yun:2022evm}.

\subsection{Source size dependence}

The size of the source can be related to a measurable quantity, the charged-particle pseudorapidity density at 
mid-rapidity, $[dN_{ch}/d\eta \,(|\eta|<0.5)$, which, in turn, can be related to 
the freeze-out temperature by means of the empirical formula~\cite{LeRoux:2021adw,QCDSR:21,Abreu:2023aqy}:
\begin{align}
    T_F=T_{F0} \ e^{-b\mathcal{N}} , 
\end{align}
where $T_{F0}=132.5\MeV$, $b=0.02$, and $\mathcal{N} \equiv [dN_{ch}/d\eta \,(|\eta|<0.5)]^{1/3}$.  Assuming that 
the hadron gas undergoes a Bjorken cooling, i.e. $T=T_h  \left( \tau_h/\tau \right)^{1/3}$, then the freeze-out 
time $\tau_F$ can be written in terms of $\mathcal{N}$ as~\cite{LeRoux:2021adw,QCDSR:21,Abreu:2023aqy}:
\begin{align}
\tau_{F}=\tau_H\left( \frac{T_H}{T_{F0}}\right)^3 e^{3b\mathcal{N}}.
\label{temperaturerelation}
\end{align}
A larger source generates a bigger $\mathcal{N}$, which, from the equation above, implies a bigger 
$\tau_{F}$, i.e. a longer hadron gas phase. As a consequence, the use of Eq.~(\ref{temperaturerelation}) in 
(\ref{Eq:RatioEquation}) will give rise to a dependence of the multiplicity 
$N_h$ on the size of the source.

As shown in Ref.~\cite{Abreu:2023aqy}, empirical formulas relating 
$\mathcal{N}$ with the volume of the system ($V$), charm quark number ($N_c$) and light quark number ($N_q $) 
can also be obtained. They are 
\begin{align}
V & = 2.82 \mathcal{N}^{3} \nonumber \\
N_c & = 7.9 \times 10^{-5} \mathcal{N}^{4.8} \nonumber \\ 
N_u & = N_d = 0.37 \mathcal{N}^{3} .
\label{size1}
\end{align}
We also assume that the charm quark number and the number of $ D $ mesons ($N_D$) are proportional. Hence, the use of 
Eq.~(\ref{size1}) in (\ref{coalmod2}) to estimate the dependence of the initial conditions with  $\mathcal{N}$, 
together with (\ref{temperaturerelation}) in (\ref{Eq:RatioEquation}), generates a dependence of  $N_h$ on $\mathcal{N}$. 

\begin{figure}[!htbp]
	\centering
\includegraphics[{width=8.0cm}]{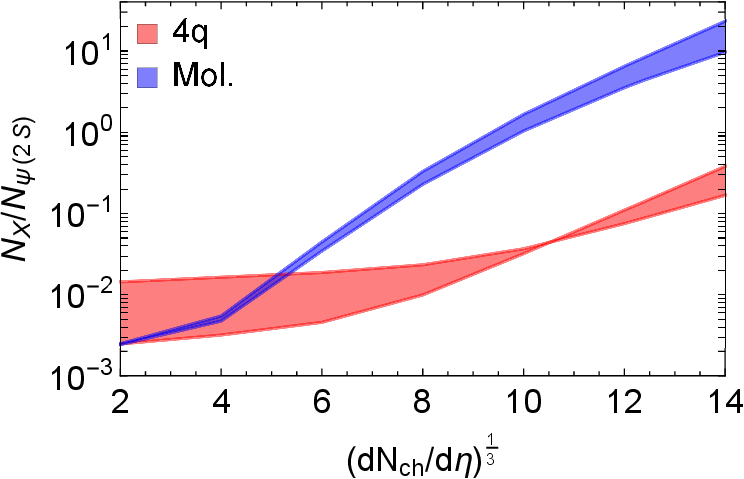} 
\caption{ Ratio $\mathcal{R}$  as a function of $\mathcal{N}$. }
\label{Fig:Abundance-dNdEta}
\end{figure}

In Fig.~\ref{Fig:Abundance-dNdEta} we observe that the ratio $\mathcal{R}$  grows with  $\mathcal{N}$. 
This is in qualitative agreement with the existing data. As mentioned in introduction, the LHCb Collaboration reported in Ref.~\cite{LHCb:2022ixc} the growth of the ratio $\mathcal{R}$  when we go from p p to 
p Pb and to Pb Pb collisions. Our results give  qualitative  support to the conjecture that the $X(3872)$ and  $\psi(2S)$ experience a different dynamics in the hadronic  medium.  Unfortunately a direct comparison is an involved task, as the data in Ref.~\cite{LHCb:2022ixc} for the p p, p Pb and Pb Pb (this last one from the CMS Collaboration) collisions refer to different energies and ranges of rapidity and transverse momentum. Interestingly, the curves of the molecular and compact tetraquark configurations converge to each other at smaller sources, but in the region of validity of this model (larger sources) the difference between these configurations remains approximately of the same order.

\subsection{Centrality and energy dependence}

Apart from its dependence on the  mass number $A$, $\mathcal{N}$ also depends on the centrality of the collision and on the center-of-mass collision energy ($\sqrt{s}$).  For Pb Pb collisions at $5.02$ TeV  the relation between $\mathcal{N}$ and the centrality (denoted as $x$, in \%) can be parametrized as ~\cite{Abreu:2023aqy}: 
\begin{eqnarray}
  \left. \frac{d N_{ch}}{d \eta}\right|_{ |\eta|<0.5} &  = & 2142.16 - 85.76
  x + 1.89 x^{2} - 0.03 x^{3} \nonumber \\ 
& &+3.67 \times 10^{-5} \ x^{4} - 2.24 \times 10^{-6} \ x^{5}   \nonumber \\ 
& &+ 5.25 \times 10^{-9} \ x^{6}, 
\label{relCN}
\end{eqnarray}
Similarly, the dependence of $\mathcal{N}$ on $\sqrt{s}$ can be parametrized as: 
\begin{eqnarray}
  \left. \frac{d N_{ch}}{d \eta}\right|_{ |\eta|<0.5} &  = & -2332.12 + 491.69 \, \log ( 220.06 + \sqrt{s} )  \nonumber \\
& &   
\label{relSN}
\end{eqnarray}
In order to determine the initial $\psi (2S)$ and $X(3872)$ multiplicities with the coalescence model in terms of the  centrality and of  $\sqrt{s}$,  we insert Eqs.~(\ref{relCN}) and~(\ref{relSN}) into Eq.~(\ref{size1}) and use these new equations in Eq.~(\ref{coalmod}).
The final yields are then obtained solving  Eq.(\ref{Eq:RatioEquation}) with these initial conditions and up to a final time  $\tau_F$ 
(which is also $\mathcal{N}$ dependent).

\begin{figure}[!htbp]
	\centering
\includegraphics[{width=8.0cm}]{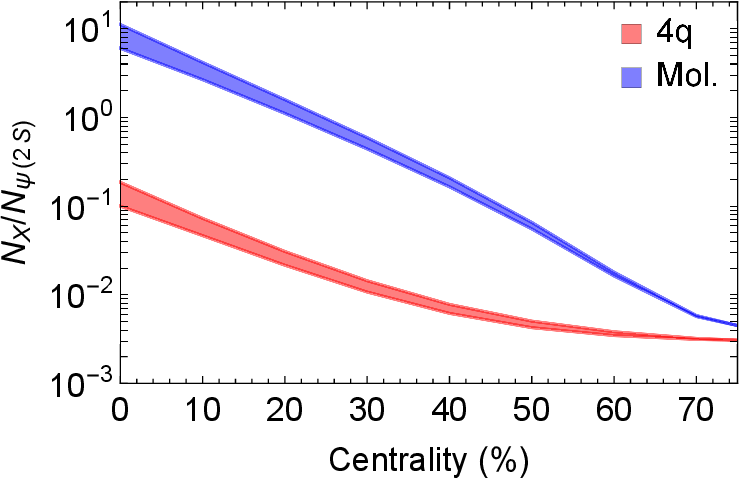} 
\caption{The ratio $\mathcal{R}$  as a function of centrality in Pb Pb 
collisions at $\sqrt{s_{NN}} =$ 5.02 TeV. }
\label{Fig:Abundance-Centrality}
\end{figure}

\begin{figure}[!htbp]
	\centering
\includegraphics[{width=8.0cm}]{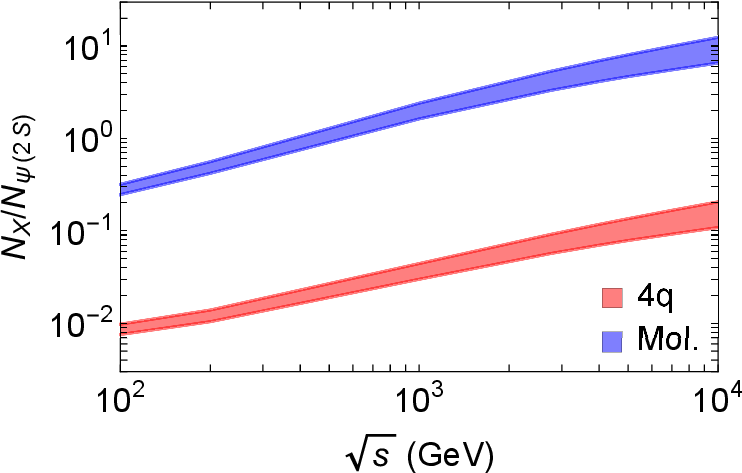} 
	\caption{ The ratio $\mathcal{R}$  as a function of energy 
$\sqrt{s}$ in central Pb Pb collisions.}
\label{Fig:Abundance-Sqrts}
\end{figure}

In Fig.~\ref{Fig:Abundance-Centrality} we show the ratio $\mathcal{R}$  as a function of the centrality. The $X(3872)$ final yield decreases 
faster than the one of $\psi(2S)$ as we move from central to peripheral collisions, and therefore the ratio becomes smaller. 
Also, the curves for the molecular and compact tetraquark configurations converge to similar values for more peripheral collisions.

The ratio  $\mathcal{R}$  as a function of the energy $\sqrt{s}$ in central Pb Pb collisions is presented in Fig.~\ref{Fig:Abundance-Sqrts}. The $X(3872)$ final yield presents an enhancement compared to that of $\psi(2S)$, resulting in a bigger ratio as $\sqrt{s}$ increases; the difference between the curves of the molecular and compact tetraquark configurations remain of the same order at different energies. 

Taking together the results in Figs. \ref{Fig:Abundance-Time}, \ref{Fig:Abundance-dNdEta}, \ref{Fig:Abundance-Centrality} and 
\ref{Fig:Abundance-Sqrts} we conclude that in central Pb Pb collisions 
at $\sqrt{s_{NN}} = 5.02$ TeV the ratio $\mathcal{R}$  for 
minimum bias events is:
\begin{eqnarray}
  \frac{N_{X}}{N_{\psi(2S)}} &  \simeq & 5   \hskip1.25cm  \mbox{for molecules} \nonumber \\
  \frac{N_{X}}{N_{\psi(2S)}} &  \simeq & 0.1 \hskip1cm     \mbox{for tetraquarks}
\label{ratfin}
\end{eqnarray}
This is the main result of our work and it is our prediction for future measurements 
at the run 3 of the ALICE Collaboration, where it will be possible to measure 
$X(3872)$ and $\psi(2S)$ at low transverse momentum, i.e. $2 < p_T < 8$ GeV. 
With data taken in this $p_T$ range it will be possible, for the first time, to 
investigate the medium effects on $X(3872)$, on $\psi(2S)$ and on their ratio and 
check our predictions. Moreover, in view of the numbers above we may have a good   
chance to discriminate between molecules and tetraquarks. From the figures we also 
conclude that is easier to identify the different configurations in larger systems   
and in more central collisions. Changing the collision energy is not so relevant for 
this purpose. 

We close this section mentioning that, apart from our work, there are other papers  
~\cite{Yun:2022evm,Andronic:2019} where the $X(3872)$ and $\psi(2S)$  $p_T$ 
distributions have been calculated with hydrodynamical and statistical models.

\section{Concluding remarks}

In this work we have studied the $X(3872)$ to $\psi(2S)$ yield ratio in heavy-ion collisions, taking into account the interactions of the $\psi (2S) $ and $ X(3872)$ states with a hadron gas made of light mesons. To this end, the thermally-averaged cross sections for the production and absorption of the $\psi(2S)$ have been evaluated for the first time by using effective Lagrangians. These cross sections, together with the thermally-averaged cross sections for the  $X(3872)$-production and absorption analyzed in previous works,  have been employed in the rate equation to determine the time evolution of  $N_{\psi(2S)}$, $N_{X}$ and of $ N_{X} / N_{\psi(2S)}$. The coalescence model has been used to compute the initial multiplicities, with the $X(3872)$ being treated both as a  molecular bound state and as a compact tetraquark. Our results indicate that the ratio is strongly affected by the combined effects of hadronic interactions and hydrodynamical expansion, and that the molecular configuration has a final value bigger than one, while the compact tetraquark gives a smaller ratio by a factor of about 50.

Other interesting finding is that the ratio $\mathcal{R}$  grows with the size of the source, which is in qualitative agreement with the data from LHCb and CMS Collaborations ~\cite{LHCb:2022ixc}, although in our case this growth has a different origin. 

We believe that this study can be seen as a theoretical support to the idea that the $X(3872)$ and the conventional charmonium $\psi(2S)$ 
have a different dynamics in a hadronic medium. 
 
\begin{acknowledgements}

This work was supported by the Brazilian agencies CNPq, FAPESP and CNPq/FAPERJ (Project INCT-F\'isica Nuclear e Aplica\c{c}\~oes - Contract No. 464898/2014-5). The authors are deeply grateful to R. Rapp, who encouraged us to study this subject and with whom we had so many fruitful discussions. We would like to thank Su-Houng Lee and Marcelo Munhoz for useful discussions. 

\end{acknowledgements}



\appendix
\section{Amplitudes}
\label{Ampl}

Here the expressions for the contributions to the amplitudes in Eq.~(\ref{eq:Amplitudes}), associated to the reactions depicted in Figs.~\ref{DIAG1} and.~\ref{DIAG2}, are given: 

\begin{eqnarray}
	\nonumber \mathcal{M}^{ (a) } &=& g_{\psi D D^{*}} g_{\pi D D^{*}}  \varepsilon_{1}^{\mu} \frac{1}{t – m_{D^{*}}^{2}} \epsilon_{\mu \beta \lambda \rho}
	\nonumber \\
	& &\times \left( - g^{\alpha \beta} + \frac{ (p_{1} – p_{3})^{\alpha} (p_{1} – p_{3})^{\beta} }{m_{D^{*}}^{2}} \right) p_{2\alpha} p_{1}^{\lambda} p_{3}^{\rho} ,
\end{eqnarray}

\begin{eqnarray}
	\nonumber \mathcal{M}^{ (b) } &=& g_{\psi D D^{*}} g_{\pi D D^{*}}  \varepsilon_{1}^{\mu} \frac{1}{u – m_{D^{*}}^{2}} \epsilon_{\mu \beta \lambda \rho}
	\nonumber \\
	& &\times \left( - g^{\alpha \beta} + \frac{ (p_{1} – p_{4})^{\alpha} (p_{1} – p_{4})^{\beta} }{m_{D^{*}}^{2}} \right) p_{2\alpha} p_{1}^{\lambda} p_{4}^{\rho} ,
\end{eqnarray}

\begin{eqnarray}
	\nonumber \mathcal{M}^{ (c) } &=& 2 g_{\psi D D} g_{\pi D D^{*}}  \varepsilon_{1}^{\mu} \varepsilon_{4}^{*\nu}  \frac{1}{t – m_{D}^{2}} p_{2\nu} p_{3\mu}, 
\end{eqnarray}

\begin{eqnarray}
	\nonumber \mathcal{M}^{ (d) } &=& \frac{1}{2} g_{\psi D^{*} D^{*}} g_{\pi D D^{*}}  \varepsilon_{1}^{\mu} \varepsilon_{4}^{*\nu}  \frac{1}{u – m_{D^{*}}^{2}} 
	\nonumber \\
	\nonumber & &\times \left( -g^{\alpha \beta} + \frac{ (p_{1} – p_{4})^{\alpha} (p_{1} – p_{4})^{\beta} }{ m_{D^{*}}^{2} } \right) (p_{2} + p_{3})_{\alpha}
	\nonumber \\
	& &\times  [2p_{4\mu} g_{\nu \beta}  - (p_{1} + p_{4})_{\beta} g_{\mu \nu} + 2p_{1\nu} g_{\mu \beta} ],
\end{eqnarray}

\begin{eqnarray}
	\nonumber \mathcal{M}^{ (e) } &=& g_{\psi D D^{*}} g_{\pi D^{*} D^{*}}  \varepsilon_{1}^{\mu} \varepsilon_{4}^{*\nu}  \frac{1}{t – m_{D^{*}}^{2}} \epsilon_{\mu \gamma \delta \beta} \epsilon_{\nu \lambda \rho \alpha}
 	\nonumber \\
	& &\times \left( -g^{\alpha \beta} + \frac{ (p_{1} – p_{3})^{\alpha} (p_{1} – p_{3})^{\beta} }{ m_{D^{*}}^{2} } \right) p_{}^{\gamma} p_{2}^{\rho} p_{3}^{\delta} p_{4}^{\lambda} ,
\end{eqnarray}

\begin{eqnarray}
	\nonumber \mathcal{M}^{ (f) } &=& g_{\psi D D^{*}} g_{\pi D D^{*}} \varepsilon_{1}^{\mu} \varepsilon_{3}^{*\lambda} \varepsilon_{4}^{*\nu} \frac{1}{t – m_{D}^{2}} \epsilon_{\mu \lambda \gamma \delta} p_{1}^{\gamma} p_{2\nu} p_{3}^{\delta},
\end{eqnarray} 

\begin{eqnarray}
	\nonumber \mathcal{M}^{ (g) } &=& - g_{\psi D D^{*}} g_{\pi D D^{*}} \varepsilon_{1}^{\mu} \varepsilon_{3}^{*\lambda} \varepsilon_{4}^{*\nu} \frac{1}{u – m_{D}^{2}} \epsilon_{\mu \nu \gamma \delta} p_{1}^{\gamma} p_{2\lambda} p_{4}^{\delta},
\end{eqnarray}

\begin{eqnarray}
	\nonumber \mathcal{M}^{ (h) } &=& g_{\psi D^{*} D^{*}} g_{\pi D^{*} D^{*}} \varepsilon_{1}^{\mu} \varepsilon_{3}^{*\lambda} \varepsilon_{4}^{*\nu} \frac{1}{t – m_{D^{*}}^{2}} \epsilon_{\nu \gamma \delta \alpha}
	\nonumber \\
	\nonumber & &\times  \left( - g^{\alpha \beta} + \frac{ (p_{1} – p_{3})^{\alpha} (p_{1} – p_{3})^{\beta} }{m_{D^{*}}^{2}} \right)  p_{4}^{\gamma} p_{2}^{\delta}
	\nonumber \\
	& &\times  [2p_{3\mu} g_{\beta \lambda} – (p_{1} + p_{3})_{\beta} g_{\mu \lambda}  + 2 p_{1\lambda} g_{\mu \beta}],
\end{eqnarray}

\begin{eqnarray}
	\nonumber \mathcal{M}^{ (i) } &=& g_{\psi D^{*} D^{*}} g_{\pi D^{*} D^{*}} \varepsilon_{1}^{\mu} \varepsilon_{3}^{*\lambda} \varepsilon_{4}^{*\nu} \frac{1}{u – m_{D^{*}}^{2}} \epsilon_{\lambda \gamma \delta \alpha}
	\nonumber \\
	\nonumber & &\times  \left( - g^{\alpha \beta} + \frac{ (p_{1} – p_{4})^{\alpha} (p_{1} – p_{4})^{\beta} }{m_{D^{*}}^{2}} \right) p_{2}^{\gamma} p_{3}^{\delta}
	\nonumber \\
	& &\times  [2p_{4\mu} g_{\beta \nu} – (p_{1} + p_{4})_{\beta} g_{\mu \nu}  + 2 p_{1nu} g_{\mu\beta}],
\end{eqnarray}

\begin{eqnarray}
	\nonumber \mathcal{M}^{ (j) }  &=& 4 g_{\psi D D} g_{\rho D D} \varepsilon_{1}^{\mu} \varepsilon_{2}^{\nu} \frac{1}{t – m_{D}^{2}} p_{3\mu} p_{4\nu},
\end{eqnarray}

\begin{eqnarray}
	\nonumber \mathcal{M}^{ (k) } &=& 4 g_{\psi D D} g_{\rho D D} \varepsilon_{1}^{\mu} \varepsilon_{2}^{\nu} \frac{1}{u – m_{D}^{2}} p_{3\nu} p_{4\mu},
\end{eqnarray}

\begin{eqnarray}
	\nonumber \mathcal{M}^{ (l) } &=& g_{\psi D D^{*}} g_{\rho D D^{*}} \varepsilon_{1}^{\mu} \varepsilon_{2}^{\nu} \frac{1}{t – m_{D^{*}}^{2}} \epsilon_{\nu \gamma \delta \alpha} \epsilon_{\mu \lambda \rho \beta} 
	\nonumber \\
	& &\times  \left( – g^{\alpha \beta } + \frac{ (p_{1} – p_{3})^{\alpha} (p_{1} – p_{3})^{\beta} }{ m_{D^{*}}^{2} } \right)  p_{1}^{\lambda} p_{2}^{\gamma} p_{3}^{\rho} p_{4}^{\delta},
\end{eqnarray}

\begin{eqnarray}
	\nonumber \mathcal{M}^{ (m) } &=& - g_{\psi D D^{*}} g_{\rho D D^{*}} \varepsilon_{1}^{\mu} \varepsilon_{2}^{\nu} \frac{1}{u – m_{D^{*}}^{2}} \epsilon_{\nu \gamma \delta \alpha} \epsilon_{\mu \lambda \rho \beta} 
	\nonumber \\
	& &\times  \left( – g^{\alpha \beta} + \frac{ (p_{1} – p_{4})^{\alpha} (p_{1} – p_{4})^{\beta} }{ m_{D^{*}}^{2} } \right)  p_{1}^{\lambda} p_{2}^{\delta} p_{3}^{\gamma} p_{4}^{\rho},
\end{eqnarray}

\begin{eqnarray}
	\nonumber \mathcal{M}^{ (n) } &=& - 2 g_{\psi D D} g_{\rho D D^{*}} \varepsilon_{1}^{\mu} \varepsilon_{2}^{\nu} \varepsilon_{4}^{*\lambda} \frac{1}{t – m_{D}^{2}} \epsilon_{\nu \lambda \gamma \delta} p_{2}^{\delta} p_{4}^{\gamma} p_{3\mu},
\end{eqnarray} 

\begin{eqnarray}
	\nonumber \mathcal{M}^{ (o) } &=& 2 g_{\psi D D} g_{\rho D D^{*}} \varepsilon_{1}^{\mu} \varepsilon_{2}^{\nu} \varepsilon_{4}^{*\lambda} \frac{1}{u – m_{D}^{2}} \epsilon_{\lambda \mu \gamma \delta} p_{1}^{\gamma} p_{4}^{\delta} p_{3\nu},
\end{eqnarray}

\begin{eqnarray}
	\nonumber \mathcal{M}^{ (p) } &=&  g_{\psi D D^{*}} g_{\rho D^{*} D^{*}} \varepsilon_{1}^{\mu} \varepsilon_{2}^{\nu} \varepsilon_{4}^{*\lambda} \frac{1}{t – m_{D^{*}}^{2}} \epsilon_{\mu \gamma \delta \beta}
	\nonumber \\
	\nonumber & &\times  \left( -g^{\alpha \beta } + \frac{(p_{1} - p_{3})^{\alpha } (p_{1}-{3})^{\beta }}{m_{D^{*}}^{2}} \right) p_{1}^{\gamma } p_{3}^{\delta }
	\nonumber \\
	& &\times  [2p_{4\nu} g_{\lambda \alpha} – (p_{2} + p_{4})_{\alpha} g_{\nu \lambda} + 2p_{2\lambda} g_{\nu \alpha} ],
\end{eqnarray}

\begin{eqnarray}
	\nonumber \mathcal{M}^{ (q) } &=&  g_{\psi D^{*} D^{*}} g_{\rho D D^{*}} \varepsilon_{1}^{\mu} \varepsilon_{2}^{\nu} \varepsilon_{4}^{*\lambda} \frac{1}{u – m_{D^{*}}^{2}} \epsilon_{\nu \gamma \delta \beta}
	\nonumber \\
	\nonumber & &\times  \left( – g^{\alpha \beta} + \frac{ (p_{1} – p_{4})^{\alpha} (p_{1} – p_{4})^{\beta} }{m_{D^{*}}^{2}} \right) p_{2}^{\delta} p_{3}^{\gamma}
	\nonumber \\
	& &\times  [2p_{4\mu} g_{\beta \lambda} – (p_{1} + p_{4})_{\beta} g_{\mu \lambda} + 2p_{1\lambda} g_{\mu \beta} ],
\end{eqnarray} 	\nonumber \\

\begin{eqnarray}
	\nonumber \mathcal{M}^{ (r) } &=& - g_{\psi D D^{*}} g_{\rho D D^{*}}  \varepsilon_{1}^{\mu} \varepsilon_{2}^{\nu} \varepsilon_{3}^{*\rho} \varepsilon_{4}^{*\lambda} \frac{1}{t – m_{D}^{2}} 
	\nonumber \\
	& & \times \epsilon_{\nu \lambda \gamma \delta} \epsilon_{\mu \rho \alpha \beta} p_{1}^{\alpha} p_{2}^{\delta} p_{3}^{\delta} p_{4}^{\gamma} ,
\end{eqnarray}

\begin{eqnarray}
	\nonumber \mathcal{M}^{ (s) } &=& g_{\psi D D^{*}} g_{\rho D D^{*}}  \varepsilon_{1}^{\mu} \varepsilon_{2}^{\nu} \varepsilon_{3}^{*\rho} \varepsilon_{4}^{*\lambda} \frac{1}{u – m_{D}^{2}} 
	\nonumber \\
	& & \times \epsilon_{\nu \rho \gamma \delta} \epsilon_{\mu \lambda \alpha \beta} p_{1}^{\alpha} p_{2}^{\gamma} p_{3}^{\delta} p_{4}^{\beta} ,
\end{eqnarray}

\begin{eqnarray}
	\nonumber \mathcal{M}^{ (t) } &=& g_{\psi D^{*} D^{*}} g_{\rho D^{*} D^{*}}  \varepsilon_{1}^{\mu} \varepsilon_{2}^{\nu} \varepsilon_{3}^{*\rho} \varepsilon_{4}^{*\lambda} \frac{1}{t – m_{D^{*}}^{2}} 
	\nonumber \\
	\nonumber & &\times  \left( - g^{\alpha \beta} + \frac{ (p_{1} – p_{3})^{\alpha} (p_{1} – p_{3})^{\beta} }{m_{D^{*}}^{2}} \right) 
	\nonumber \\
	\nonumber & &\times  [2p_{4\nu} g_{\alpha \lambda} – (p_{2} + p_{4})_{\alpha} g_{\nu \lambda} + 2p_{2\lambda} g_{\nu \alpha} ]
	\nonumber \\
	& &\times  [2p_{3\mu} g_{\beta \rho} – (p_{1} + p_{3})_{\beta} g_{\mu \rho} + 2p_{1\rho} g_{\mu \beta} ],
\end{eqnarray}

\begin{eqnarray}
	\nonumber \mathcal{M}^{ (u) } &=& g_{\psi D^{*} D^{*}} g_{\rho D^{*} D^{*}}  \varepsilon_{1}^{\mu} \varepsilon_{2}^{\nu} \varepsilon_{3}^{*\rho} \varepsilon_{4}^{*\lambda} \frac{1}{u – m_{D^{*}}^{2}} 
	\nonumber \\
	\nonumber & &\times  \left( - g^{\alpha \beta} + \frac{ (p_{1} – p_{4})^{\alpha} (p_{1} – p_{4})^{\beta} }{m_{D^{*}}^{2}} \right) 
	\nonumber \\
	\nonumber & &\times  [2p_{3\nu} g_{\alpha \rho} – (p_{2} + p_{3})_{\alpha} g_{\nu \rho} + 2p_{2\rho} g_{\nu \alpha} ]
	\nonumber \\
	& &\times  [ 2p_{4\mu} g_{\beta \lambda} – (p_{1} + p_{4})_{\beta} g_{\mu \lambda} + 2p_{1\lambda} g_{\mu \beta} ],
\end{eqnarray}
where $p_1$ and $p_2$ are the momenta of initial state particles, 
while $p_3$ and $p_4$ are those of final state particles; $s,t,u$ are the Mandelstam variables: $s = (p_1 +p_2)^2, t = (p_1 - p_3)^2,$ and $u = (p_1-p_4)^2$;  and  $\epsilon_{\textbf{i}}  ^{(*)} \equiv \epsilon ^{(*)}(p_i) $ is the polarization vector.


\end{document}